\documentstyle[12pt]{article}

\renewcommand{\thefootnote}{\fnsymbol{footnote}}
\renewcommand{\theenumi}{(\roman{enumi})}
\renewcommand{\thefigure}{\arabic{figure}}
\def\beq{\begin{equation}}
\def\eeq{\end{equation}}
\def\bea{\begin{eqnarray}}
\def\eea{\end{eqnarray}}
\def\nn{\nonumber}
\def\dfrac{\displaystyle\frac}
\def\bs{\bigskip}

\def\PR#1#2#3{Phys.~Rev. {\bf #1}, #2  (#3)}
\def\PRL#1#2#3{Phys.~Pev.~Lett. {\bf #1}, #2 (#3)}
\def\PL#1#2#3{Phys.~Lett. {\bf #1}, #2 (#3)}

\def\PTP#1#2#3{Prog.~Theor.~Phys. {\bf #1}, #2 (#3)}

\def\eqref#1{eq.(\ref{eqn:#1})}
\def\Eqref#1{Equation (\ref{eqn:#1})}
\def\eqlab#1{\label{eqn:#1}}

\def\bmaT{\left(\begin{array}{ccc}}
\def\emaT{\end{array}\right)}

\def\vev#1{\langle #1 \rangle}

\def\l{\left}
\def\r{\right}
\def\eg{{\it e.g.}}

\def\Su{\rm u}
\def\Sd{\rm d}
\def\Sud{\rm u,d}

\def\he{{y^{\rm e}}}

\begin{document}
\renewcommand{\thefootnote}{\fnsymbol{footnote}}
\renewcommand{\theenumi}{(\roman{enumi})}
\renewcommand{\thefigure}{\arabic{figure}}
\title{Energy-Scale Dependence of \\
 the Lepton-Flavor-Mixing Matrix}
\author{
{N. Haba$^{1,2}$}\thanks{E-mail address:haba@pacific.mps.ohio-state.edu}
{, Y. Matsui$^3$}\thanks{E-mail address:matsui@eken.phys.nagoya-u.ac.jp} 
{, N. Okamura$^4$}\thanks{E-mail address:naotoshi.okamura@kek.jp} 
{ and M. Sugiura$^3$}\thanks{E-mail address:sugiura@eken.phys.nagoya-u.ac.jp} 
\\
\\
\\
{\small \it $^1$Department of Physics, The Ohio State University,}
{\small \it Columbus, Ohio 43210, USA}\\
{\small \it $^2$Faculty of Engineering, Mie University,}
{\small \it Tsu Mie 514-8507, Japan}\\
{\small \it $^3$Department of Physics, Nagoya University,}
{\small \it Nagoya, 464-8602, Japan}\\
{\small \it $^4$Theory Group, KEK, Tsukuba Ibaraki 305-0801, Japan}}
\date{}
\maketitle
\vspace{-12.0cm}
\begin{flushright}
hep-ph/9904292\\
OHSTPY-HEP-T-99-010\\
DPNU-99-11\\
KEK-TH-620
\end{flushright}
\vspace{12.0cm}
\vspace{-2.5cm}
%
%
\begin{abstract}

We study an energy-scale dependence of the lepton-flavor-mixing matrix
in the minimal supersymmetric standard model with 
the effective dimension-five operators
which give the masses of neutrinos.
We analyze the renormalization group equations of $\kappa_{ij}$s
which are coefficients of these effective operators under 
the approximation to neglect the corrections of $O(\kappa^2)$. 
As a consequence,
we find that all phases in $\kappa$ do not depend on the energy-scale, 
and that only $n_g^{}-1$ ($n_g^{}$: generation number)
real independent parameters in the lepton-flavor-mixing
matrix depend on the energy-scale.

\end{abstract}


\newpage

%
%


 Recent neutrino experiments suggest the 
existence of the flavor mixing 
in the lepton sector
\cite{solar3}-\cite{LSND}. 
 Studies of the lepton-flavor-mixing matrix, 
which is called \cite{no6} Maki-Nakagawa-Sakata (MNS) matrix \cite{MNS},
open a new era of the flavor physics. 
 We can predict 
the lepton-flavor-violating interactions such as
$\mu \rightarrow e \gamma$ from the MNS matrix. 
 When we consider that the lepton-flavor-violating interactions are 
related to the new physics at a high energy-scale,
it is important to analyze the energy-scale dependence
of the MNS matrix
in order to obtain information of the new physics.

 In this letter we study the energy-scale dependence of the MNS matrix
in the minimal supersymmetric standard model (MSSM) with
effective dimension-five operators
which give masses of neutrinos.
 In this model the superpotential of lepton-Higgs interaction terms is
\beq
{\cal W} =
 {\he}_{ij} H_{\Sd} L_i {\bar E_{j}}
 - \dfrac{1}{2} \kappa^{}_{ij} (H_{\Su} L_i)(H_{\Su} L_j) \,.
\eqlab{superpot}
\eeq
Here the indices $i,j$ $(=1 \sim n_g^{})$ stand for the generation number.
$L_i$ and ${\bar E_{i}}$ are $i$-th generation 
lepton doublet and right-handed charged lepton,
$H_{\Sud}$ are the Higgs doublets
which give Dirac masses to the up- and down-type fermions, 
respectively.
 The coefficients matrix ($\kappa$)
of the effective dimension-five operators
which is an $n_g^{}\times n_g^{}$ 
complex and symmetric matrix,
gives the neutrino Majorana mass matrix.
When we take the diagonal base of
the charged lepton Yukawa coupling $\he$,
$\kappa$ is diagonalized by the MNS matrix.
 All the elements of $\kappa$ are naturally small
if they are generated effectively by the new physics 
at a high energy-scale $M$.  
 One of the most reliable candidate 
is so-called see-saw mechanism \cite{seesaw}, 
where the small $\kappa$ of $O(1/M)$ is generated 
by the heavy right-handed neutrinos with Majorana masses of $O(M)$. 

%

{\bs}
 Now let us consider the renormalization of $\kappa$.
 The wave function renormalization of $L_i$ is given
by $L_i^{(0)} = Z_{ij}^{-1/2} L_j$
and that of the Higgs doublet is given by
$H_{\Su}^{(0)} = Z_H^{-1/2} H_{\Su}$.
Then the renormalization of $\kappa^{}_{ij}$ is written as
\beq
\kappa^{(0)}_{ij} =
\l(Z^{-1/2}_{ik} Z^{-1/2}_{jl} Z^{-1}_H\r)
\kappa^{}_{kl}\,.
\eqlab{Renorm0}
\eeq
 Here we adopt the approximation to neglect loop corrections of
$O(\kappa^2)$,
which are sufficiently
small according to the tiny neutrino masses.
It corresponds to 
taking the Feynman diagrams 
in which $\kappa$ appears only once. 
 If $\kappa$ is induced by the see-saw mechanism,
this approximation is consistent with neglecting
terms of $O(1/M^2)$ in the see-saw mechanism.
 Under this approximation
$Z_{ik}$ becomes diagonal as
$Z_{ik}=Z_i \delta_{ik} + O(\kappa^2)$
because there are no lepton-flavor-mixing terms except $\kappa$.
Therefore \eqref{Renorm0} becomes simple as
\beq
  \kappa_{ij}^{(0)} =
  \l(Z_i^{-1/2}Z_j^{-1/2}Z_H^{-1}\r) \kappa^{}_{ij}\,.
  \eqlab{Renorm1}
\eeq
\Eqref{Renorm1} leads to the RGE of $\kappa^{}_{ij}$ as 
\beq
\dfrac{d}{dt} \kappa^{}_{ij} = 
\l( \gamma_i^{} + \gamma_j^{} + 2 \gamma_H^{} \r)
\kappa^{}_{ij}\,,
\eqlab{kappa_g_1}
\eeq
where $t$ is the scaling parameter
which is related to the renormalization scale 
$\mu$ as $t=\ln\mu$. 
$\gamma_i$ and $\gamma_H$ are defined as
\beq
  \gamma_i^{} = \dfrac12 {\dfrac{d}{dt}} \ln Z_i^{}\,,
  \mbox{\quad}
  \gamma_H^{} = \dfrac12 {\dfrac{d}{dt}} \ln Z_H^{}\,.
\eqlab{anom_dim}
\eeq

{\bs}

By using \eqref{kappa_g_1},
 we can obtain the following two consequences:\\
(1){\it
All phases in $\kappa$ do not 
depend on the energy-scale.
}
 By using the notation 
$\kappa^{}_{ij} \equiv |\kappa^{}_{ij}| e^{i \varphi_{ij}}$,
\eqref{kappa_g_1}
is rewritten as 
\bea
\dfrac{d}{dt} \ln{\kappa^{}_{ij}}
&=& \dfrac{d}{dt} \ln{|\kappa^{}_{ij}|}+ i \dfrac{d}{dt}\varphi_{ij} \nn 
\\
&=& \l( \gamma_i^{} + \gamma_j^{} + 2 \gamma_H^{} \r)\,.
\eqlab{kappa_g_2}
\eea
Since $\gamma_i$, $\gamma_j$ and $\gamma_H$ are real,
\eqref{kappa_g_2} means
\beq
\dfrac{d}{dt}\varphi_{ij} = 0\,.
\eeq
Therefore we can conclude that
the arguments of all the elements of $\kappa$ are not changed
by RG evolutions.
 We should notice that this result does not necessarily mean 
that phases of the MNS matrix are independent of 
the energy-scale as we will see later. 

\noindent
(2){\it
 Only $n_g^{}-1$ real independent parameters in the MNS matrix
depend on the energy-scale.
}
The combinations of $\kappa$'s elements,
\beq
 c_{ij}^2=\dfrac{\kappa_{ij}^2}{\kappa_{ii}^{} \kappa_{jj}^{}}\,,
\eeq
 are the energy-scale independent quantities because 
\bea
\dfrac{d}{dt} \ln\l(\dfrac{\kappa_{ij}^2}{\kappa^{}_{ii} \kappa^{}_{jj}}\r)
&=& 2 \dfrac{d}{dt} \ln \kappa^{}_{ij}
    - \dfrac{d}{dt} \ln \kappa^{}_{ii}
    - \dfrac{d}{dt} \ln \kappa^{}_{jj} \nn \\
&=& 2\l(\gamma_i^{} + \gamma_j^{} + 2 \gamma_H^{} \r)
    -\l(2\gamma_i^{} + 2 \gamma_H^{} \r)
    -\l(2\gamma_j^{} + 2 \gamma_H^{} \r) \nn \\
&=& 0\,.
\eqlab{cij}
\eea
 Since the off-diagonal 
elements of the $\kappa^{}_{ij}$ $(i \neq j)$
are given by 
\beq
\kappa^{}_{ij}=c_{ij} \sqrt{\kappa^{}_{ii} \kappa^{}_{jj}}
\mbox{\quad} (i \neq j)\,,
\eqlab{kappa_ij}
\eeq
 their energy-scale dependence 
can be completely determined
by the diagonal elements ${\kappa^{}_{ii}}$. 
 The diagonal elements $\kappa^{}_{ii}$ can always be 
taken to be real by rephasing neutrino fields, and 
they never become complex by the RGE in \eqref{kappa_g_1}.
 The RGE of $\kappa$ can be written by only
$n_g^{}$ equations.
 The diagonal form of $\he$ is held at any energy-scale
because there is no lepton-flavor-violating correction
to the RGE of $\he$ up to $O(\kappa)$.
 Since the overall factor of the $\kappa$'s elements
is nothing to do with the MNS matrix,
the energy-scale dependence of the MNS matrix can be 
determined by $n_g^{}-1$ real independent parameters.

%

{\bs}

 Let us show an example
for the case of three generations.
 The matrix $\kappa$ is parameterized as
\beq
\kappa = \kappa^{}_{33}
\bmaT
r^{}_1 & c^{}_{12} \sqrt{r_1 r_2} & c^{}_{13} \sqrt{r_1} \\
c^{}_{12} \sqrt{r_1 r_2} & r_2 & c^{}_{23} \sqrt{r_2} \\
c^{}_{13} \sqrt{r_1} & c^{}_{23} \sqrt{r_2} & 1
\emaT\,,
\eqlab{def_kappa}
\eeq 
where 
\beq
r_i \equiv \dfrac{\kappa^{}_{ii}}{\kappa^{}_{33}}\,, 
\mbox{\quad} (i=1,2)\,.
\eqlab{def_ki}
\eeq
 The complex parameters, $c^{}_{ij}$ are energy-scale independent. 
 There are nine degrees of freedom, which are 
three complex constants $c_{ij}^{}\;(i\neq j)$ and 
three energy-scale dependent real parameters
$r_1$, $r_2$ and $\kappa^{}_{33}$. 
 Since $\he$ has diagonal form at any energy-scale and 
$\kappa^{}_{33}$ is nothing to do with
the MNS matrix, only two parameters, 
$r_1$ and $r_2$, 
are energy-scale dependent parameters in the MNS matrix.

 Here we roughly estimate 
the energy-scale dependence of the $r_i$ in \eqref{def_ki}
by using the one-loop RGEs in the MSSM \cite{nue-RGE,no5}.
We can easily show the RGE of $r_i$ is given by 
\beq
\dfrac{d}{dt} \ln r_i 
=\dfrac{d}{dt} \ln \dfrac{\kappa^{}_{ii}}{\kappa^{}_{33}} 
= -\dfrac{1}{8\pi^2}\l(y_{\tau}^2-y_{i}^2 \r)\,,
\mbox{\quad}
(i=1,2)\,,
\eqlab{RGE_kr}
\eeq
 where $y_\tau$ and $y_i$ are Yukawa couplings of $\tau$ and $i$-th
generation charged lepton, respectively.
 Neglecting the energy-scale dependence of $y_{\tau}$, 
the magnitude of the right-hand-side in \eqref{RGE_kr} is 
roughly given by 
$ y_{\tau}^2(m_Z^{})/{8\pi^2} 
= O(10^{-6}) / {\cos^2 \beta}$,
where $m_Z^{}$ stands for the weak scale
and $\tan\beta = \vev{H_{\Su}}/\vev{H_{\Sd}}$.
 It means that $r_i$s are not sensitive
to the energy-scale.
 We should stress here that this fact does not necessarily
result in the tiny energy dependence of the MNS matrix.
 We can explicitly see the significant RGE corrections
of the MNS matrix in some situations \cite{nue-RGE,no5}.
 In ref.~\cite{no5}, the drastic change of the MNS matrix
by the RGE was obtained when neutrinos of 
the second and third generations have masses
of $O(\mbox{\rm eV})$ with $\delta m_{23}^2 \simeq 3 \times
10^{-3}$ (eV$^2$) \cite{SK1}.
 This situation corresponds to the case of 
$r_1 \sim |c_{12}| \sim 0$, $r_2 \sim 1$,
and $|c_{23}| \ll 1$ in eq.(11), 
where the slight change of $r_2$ induces 
the maximal mixing of the second and third generations 
in the MNS matrix.

%

{\bs}

 In this letter
we studied an energy-scale dependence of the MNS matrix 
in the MSSM with the effective dimension-five operator.
 The coefficient of the dimension-five operator $\kappa$ is small enough
to neglect corrections of $O(\kappa^2)$ in RGEs.
 Under this approximation we found that
all phases in $\kappa$ do not depend on the energy-scale, and
that only $n_g^{}-1$ real independent parameters in the MNS matrix
depend on the energy-scale.
 Our consequences imply that there must be
$(n_g^{}-1)^2 = n_g^{}(n_g^{}-1) - (n_g^{} - 1)$
scale independent relations among
the MNS matrix elements
because the MNS matrix generally has 
$n_g^{}(n_g^{}-1)$ real independent parameters 
when neutrinos are Majorana fermions.
 These results can be helpful for the lepton flavor physics
and search for the new physics at the high energy-scales.

 Finally we discuss the possibility
to obtain the same consequences
in other models with the effective dimension-five operators.
 The supersymmetry (SUSY) is needed to obtain our consequences.
 Moreover,
it was necessary that the model did not have
lepton-flavor-violating terms and non-renomalizable terms
except the effective dimension-five operators.
 Thus we can obtain the same consequences in
the SUSY models which have these properties,
\eg\ the Next-MSSM.
 On the other hand we cannot directly apply our analysis
to the standard model or non-SUSY models
because non-zero vertex renormalization generates
an additional term in the right-hand-side of \eqref{kappa_g_1},
which is generally not real.
 Nevertheless, we can explicitly show that
this term is real at one-loop level \cite{nue-RGE}.
 Therefore we can obtain the same consequences
in the standard model at one-loop level.

%

{\bs}

 We would like to thank K. Hagiwara, M. Harada,
J. Hisano and A.I. Sanda for useful discussions and comments.
 NH would like to thank S. Raby and K. Tobe for
useful discussions, and 
is partially supported by DOE grant DOE/ER/01545-753. 
 NO is supported by the JSPS
Research Fellowships for Young Scientists, No. 2996.

%
%

\end{document}